\documentclass[12pt,a4paper,twoside]{article}
\usepackage{epsfig}
\usepackage{baltlat2}
\usepackage{wrapfig}
\usepackage{dcolumn}
\begin{document}
\ \
\vspace{0.5mm}
\setcounter{page}{1}
\vspace{6mm}
\titlehead{Baltic Astronomy, vol.\,15, p.}
\titleb{A cool R Coronae Borealis star Z\,UMi}

\begin{authorl}
\authorb{T\~onu Kipper}{1} and
\authorb{Valentina G. Klochkova}{2} 
\end{authorl}

\begin{addressl}
\addressb{1}{Tartu Observatory, T\~oravere, 61602, Estonia; tk@aai.ee} 

\addressb{2}{Special Astrophysical Observatory RAS, Nizhnij Arkhyz, 369167,
             Russia; valenta@sao.ru} 
\end{addressl}

\submitb{Received ..., 2006}
\begin{abstract}
The high resolution spectra of a R\,CrB type star Z\,UMi are analysed.
The atmospheric parameters of Z\,UMi are estimated $T_{\rm eff}=5250\pm250$\,K
and $\log g=0.5\pm0.3$. This places Z\,UMi among the coolest R\,CrB stars.
The hydrogen deficiency of Z\,UMi is confirmed. The abundances of other 
elements resemble those found for the minority group of R\,CrB stars. We
note very low iron abundance, [Fe/H]=$-1.85$, and an excess of
lithium,  [Li/Fe]=+1.9.
\end{abstract}
\begin{keywords}
stars: atmospheres -- stars: individual: Z\,UMi
\end{keywords}
\resthead{RCrB star Z\,UMi}{T.\,Kipper, V.\,Klochkova }

\sectionb{1}{INTRODUCTION}

Z\,UMi (an optical counterpart of an IR-source IRAS\,15060\,+\,8315)
was classified as a R Coronae Borealis star by Benson et al. (1994).
Before that Z\,UMi was thought to be a Mira variable with a possible
period of 450 or 500 days (Kholopov et al. 1985). The classification
as a R\,CrB star relied on the photometric and low resolution
spectroscopic observations. The photometry showed characteristic to R\,CrB
stars light drop by 6 magnitudes in 1992, which lasted about 300 days. Low
resolution spectra obtained during the light minimum showed only the
Na\,I\,D lines in emission and the C$_2$ Swan system bands in absorption.
No Balmer lines were detected.

The hydrogen-deficient nature of Z\,UMi was established by Gos\-wami et al.
(1997), who used the high resolution ($R\approx60\,000$) spectra of the star
during maximum light. The CN red and violet systems and the C$_2$ Swan system
bands were found in absorption. Weak or absent CH bands in Z\,UMi spectra
were considered as the indicators of hydrogen deficiency.
The high resolution ($R\approx30\,000$) spectrum of Z\,UMi during the
onset of the 1997 decline was studied by Goswami et al. (1999). 

In this paper we present the high resolution spectra of Z\,UMi during its 
maximum light. 

\sectionb{2}{OBSERVATIONS}

Our high resolution spectra were obtained with the Nasmyth Echelle
Spectrometer (Panchuk et al. 1999; Panchuk et al. 2002) of Russian 6\,m
telescope on March 10, 2004 (JD\,2453075), when the star was at maximum
light ($V\approx11.3$). The spectrograph was equipped with an image slicer
(Panchuk et al. 2003). As a detector a CCD camera with $2052\times2052$
pixels produced by the Copenhagen University Observatory was used.

The spectra cover 528--677\,nm without gaps unil 586\,nm. The spectra were
reduced using the NOAO astronomical data analysis facility IRAF. The
continuum was placed by fitting low order spline functions through the
manually indicated points in every order. The use of image slicer results
in three parallel strips of spectra in each order. These strips are
wavelength shifted. Therefore all strips were reduced separately and then
already linearized in the wavelength spectra were coadded. We checked the
accuracy of this procedure (Kipper \& Klochkova 2005) and found that the
wavelengths of the terrestial lines in the stellar spectrum were
reproduced within a few 0.001\,\AA-s. After that all spectra of the set
were coadded.
 
As measured from the Th-Ar comparison spectra the resolution is
$R\approx$\,42\,800 with FWHM of comparison lines about 7\,km\,s$^{-1}$.

\sectionb{3}{DESCRIPTION OF THE SPECTRA}

\subsectionb{3.1}{Atmospheric parameters}

No multicolor photometry is available for Z\,UMi. Goswami et al. (1997)
compared the high resolution spectra of Z\,UMi at maximum light with the
spectra of carbon star UU\,Aur (N3, C5,4) and the R1 spectral type
hydrogen-deficient (HdC) star HD\,182040. They found that the C$_2$ Swan
bands in Z\,UMi spectrum were much stronger than in HD\,182040 indicating a
lower temperature. At the same time Z\,UMi is not as cool as UU\,Aur. In
Figures 1 and 2 our spectra of Z\,UMi and HD\,182040 are compared in two
C$_2$ band-head regions. Our spectra also show much stronger C$_2$ bands
in the Z\,UMi spectrum. Asplund et al. (1997) estimated for HD\,182040
$T_{eff}$=5600\,K. This estimate took into account a correction to the
photometric temperature $T_{eff}$=$-500$\,K to reach better agreement with
the spectroscopically obtained values of other R\,CrB type stars.
Tenenbaum et al. (2005) have analyzed CO band observations in R\,CrB stars
and accepted $T_{eff} \approx$ 5000\,K for Z\,UMi on the basis of CO band
visibility. With the temperature close to 5000\,K, the mean $M_{bol}=-5\pm
1$ and the mass $\mathcal{M/M}_{\odot}=0.7\pm0.2$ of R\,CrB stars
(Asplund et al. 2000), the surface gravity will be close to $\log g =
0.5\pm0.3$. Tenenbaum et al. (2005) found that there is very little
difference in the strengths of CO absorption bands if $\log g$=0.5 or 1.5.
The current spectra are not suitable for the spectroscopic determination
of the microturbulence parameter $\xi_t$. We choose the value $\xi_t$=5\,km/s
close to the mean value of cooler R\,CrB stars. One of the parameters
necessary in constructing the atmospheric models of hydrogen-deficient
stars is the C/He abundance ratio as He is likely the most abundant
element and C\,I is thought to be the main agent of the continuous opacity.
Therefore, the element abundances from line intensities represent actually
the element-to-carbon abundance ratio (Asplund et al. 2000). The C/He ratio
for cool R CrB stars cannot be determined observationally. For hotter
R\,CrB stars and EHe stars Pandey et al. (2001) found C/He ratio (by number)
close to 0.01. We choose this value for Z\,UMi too. The input abundances of
other elements are those adopted by Asplund et al. (1997) for the R\,CrB
stars.

\subsectionb{3.2}{Model atmospheres}

The model atmospheres for this analysis were computed with a modified
version of MARCS program (Gustafsson et al. 1975) with updated opacities.
The opacities from continuum sources, molecules CO, CN, C$_2$ , HCN, C$_2$
H$_2$ and C$_3$, and the metallic lines were taken into account. The line
opacities were treated in opacity sampling approximation (J\o rgensen et al.
1992). The structure of these models fits well with the Uppsala line
blanketed models described by Asplund et al. (1997), when models with the
same parameters are compared. We started with the model 5000/0.5
(C/He=0.01). Synthesizing the spectral region with the C$_2$ Swan system
(0,1) band head we found that the computed C$_2$ rotational lines were too
strong. Rising the effective temperature to $T_{eff}$=5250\,K does not
completely remove the discrepancy. Still higher $T_{eff}$ seems to be
unacceptable due to the temperature limit set by Tenenbaum et al. (2005)
concerning the visibility of CO bands. The quite slight reduction of the
input carbon abundance to $\log \varepsilon({\rm C})$=$-2.2$ gives an acceptable
fit. Therefore the further analysis was performed with the model 5250/0.5
(C/He=0.006). We estimate the error in $T_{eff}$ about 250\,K.

\subsectionb{3.3}{Elemental abundances}

Z\,UMi shows in its spectrum many strong C$_2$ Swan system bands including
the (3,6) and (1,4) bandheads at 658.9 and 676.2\,nm which are usually not
visible in early-type R carbon stars. The lines of these bands are
naturally not included into the Bell (1976) list, which is the main used
line-list. We synthesized these bands using the line-list generated at the
Indiana University (Alexander 1991). About the spectral line-lists for
carbon stars see Kipper (2004). If these two red bands were taken into
account, the mean carbon abundance of Z\,UMi is
$\log\varepsilon(\rm C)$=$-2.04\pm0.20$. Without these bands
$\log\varepsilon(\rm C)$=$-2.10\pm0.1$. This means that for C$_2$ molecular
spectrum there is no so-called carbon problem (Asplund et al. 2000). For
C\,I lines the problem is present at the level of about $-0.5$dex as for
all other R\,CrB type stars analyzed up to now. After fixing the input
carbon abundance at the level of $-2.10$, the following nitrogen abundance
was found fitting CN rotational lines: $\log\varepsilon(\rm N)$=$-3.0\pm0.2$,
which almost coincides with the mean nitrogen abundance
$\log\varepsilon(\rm N)$=$-3.1\pm0.5$ for 20 R\,CrB stars analyzed by Asplund
et al. (2000).

Determining the abundances from individual spectral lines is extremely
ambiguous in the case of Z\,UMi. There are no lines belonging to atomic
species, which are not blended with molecular lines, and therefore no
equivalent widths could be measured and the procedures like determining
the microturbulence parameter $\xi_t$, adjusting $T_{eff}$ using abundance
versus excitation potential and log g determination using ionization
equilibrium, are not applicable. Synthesizing the spectra only the upper
limit of abundances could be found. As the lines, which could be used in
this case are strong, the errors due to the errors in the structure of
model atmosphere could not be estimated. Also, the error in the accepted
microturbulence parameter t of about 1\,km/s translates directly into
abundance errors greater than $\pm0.3$\,dex. This should be kept in mind
when using the following abundance data. The oscillator strengths of used
metallic lines in Bell's line-list were replaced with the solar ones by
Thevenin (1989, 1990). The found abundances are listed in Table\,1. The
indicated errors reflect only the scatter due to use of different lines.
The number of used lines, which is very small in most cases, is also
indicated in the column Remarks. We found that the hydrogen abundance
could not exceed 10-5 by synthesizing the H$\alpha$ line blend with CN
lines. A few (5) lines of N\,I show that $\log\varepsilon(N)\ge -3.1$.
This is in accord with the N abundance found from CN bands. The oxygen
abundance from 7 O\,I lines including [O\,I] lines at 630.02 and 636.39\,nm
is $\log\varepsilon(O)$=$-3.45\pm0.30$. The iron abundance from Fe\,I and
Fe\,II lines $\log\varepsilon(Fe)$=$-5.9\pm0.3$ is very low. The almost
coinciding abundances from Fe\,I ($-5.93$) and Fe\,II ($-5.87$) indicates
that the combination of $T_{eff}$ and $\log g$ is not too far from the
reality. As was found earlier by Goswami et al. (1997), we note that the
Li\,I line at 670.78\,nm is clearly present and we estimate
$\log\varepsilon({\rm Li})\approx 3.3$ and [Li/Fe]=1.9.
This means that Z\,UMi is indeed the ``Li-rich'' R\,CrB star.

\subsectionb{3.4}{RADIAL VELOCITY}

An important task for R\,CrB stars is a search for variability in their
radial velocity. Based on few relatively unblended metallic lines, we
determined an average value of radial velocity for the moment of star's
observation JD 2453074.6: $V_{rad}$=$-35.4\pm 1.6$\,km/s. From the
rotational lines of C$_2$ (0,1) band we found $-35.4\pm1.8$\,km/s and
(0,2) band $-35.8\pm 2.0$\,km/s. Therefore we estimate the mean
heliocentric radial velocity of Z\,UMi, $V_{rad}$=$-35.5\pm 1.7$\,km/s.
Goswami et al. (1999) estimated $V_{rad}$=$-35\pm 2$\,km/s from the narrow
emission lines in the decline spectrum of Z\,UMi. The photospheric velocity
found by Goswami et al. (1997) is $V_{rad}$=$-37\pm 2$\,km/s. Taking into
account the stated errors there are no changes in the radial velocity. The
Gaussian decomposition of the Na\,I D--lines reveals three components
presented in Table\,2. 

Two redmost and relatively sharp components could be of interstellar
origin. The strongest and bluest component shows the presence of earlier
expelled material expanding with the velocity of about 10\,km/s. The
larger blue extension and therefore the larger blueshift of the D$\rm _2$ line
compared to the D$\rm _1$ line could be caused by a heap of C$_2$ and CN lines
around 588.8\,nm.

\sectionb{4}{CONCLUSIONS}

We found that Z\,UMi is one of the coolest R\,CrB stars with its
$T_{eff}\approx 5250$\,K. The other cool R\,CrB stars S\,Aps, WX\,CrA and U\,Aqr
have $T_{eff}\approx 5000$\,K. The even cooler DY\,Per ($T_{eff}$=$3500\div4700$\,K) is a
prototype of a small group with very slow and much more symmetrical
declines than the declines of R\,CrB stars (Alcock et al. 2001). It is not
yet clear whether the DY\,Per stars are related to the R\,CrB stars. The
derived abundances are very crude estimates. In spite of that we conclude
that the chemical composition of Z\,UMi resembles that of the minority
group R\,CrB stars. Similarly to V\,CrA, VZ\,Sgr and V3795\,Sgr it is very
metal--poor. Nitrogen is overabundant as in the majority group and V\,CrA
from the minority group.

\vskip5mm
ACKNOWLEDGEMENTS.\

This research was  supported by the  Estonian Science
Foundation grant nr.~6810 (T.K.).
V.G.K. acknowledges the support from the programs of Russian Academy of Sciences
 ``Observational manifestations of evolution of chemical composition
 of stars and
Galaxy'' and ``Extended objects in Universe''. V.G.K. also acknowledges  
the support by Award No.\,RUP1--2687--NA--05 
of the U.S. Civilian Research \& Development Foundation (CRDF).

\goodbreak

\References{}
\refb
Alcock C., Allsman R.A., Alves D.R., et al. 2001, ApJ, 554, 298
\refb
Alexander D.R. 1991, private communication
\refb
Asplund M., Grevesse N., Sauval J. 2005, ASP Conf. Ser., 336, 25
\refb
Asplund M., Gustafsson B., Lambert D.L., Rao N.K 2000, A\&A, 352, 287
\refb
Asplund M., Gustafsson B., Kiselman D., Eriksson K. 1997, A\&A, 318, 521
\refb
Bell R.A. 1976, private communication,
 http://ccp7.dur.ac.uk/ccp7/DATA/lines.bell.tar.Z
\refb
Benson P.J., Clayton G.C., Garnavich P., Szkody P. 1994, AJ, 108, 247
\refb
Goswami A., Rao N.K., Lambert D.L., Gonzalez G. 1997, PASP, 109, 796
\refb
Goswami A., Rao N.K., Lambert D.L. 1999, Observatory, 119, 22
\refb
Gustafsson B., Bell R.A., Eriksson K., Norlund \AA.  1975, A\&A, 42, 407
\refb
J\o rgensen U.G., Johnson H.R., Norlund \AA.  1992, A\&A, 261, 263
\refb
Kholopov  P.N. 1985, General Catalogue of Variable Stars. Moscow. ``Nauka''.
\refb
Kipper T. 2004, Baltic Astronomy, 13, 573
\refb
Kipper T., Klochkova V.G. 2005, Baltic Astronomy, 14, 215
\refb
Panchuk V.E., Klochkova V.G., Naidenov I.D. 1999, Preprint
Spec. AO, No 135
\refb
Panchuk V.E., Piskunov N.E., Klochkova V.G., et al. 2002, Preprint
Spec. AO, No 169
\refb
Panchuk V.E., Yushkin M.V., Najdenov I.D. 2003, Preprint Spec. AO, No 179
\refb
Pandey G., Rao N.K., Lambert D.L., Jeffery C.S., Asplund M. 2001, MNRAS, 324,
937
\refb
Tenenbaum E.D., Clayton G.C., Asplund M., et al. 2005, AJ, 130, 256
\refb
Thevenin F. 1989, A\&AS, 77, 137
\refb
Thevenin F. 1990, A\&AS, 82, 179
\refb
Wiese W.L., Fuhr J.R., Deters T.M. 1996, J. Chem. Ref. Data, Mono. 7

\newpage

\begin{figure}
\vskip2mm
\centerline{{\psfig{figure=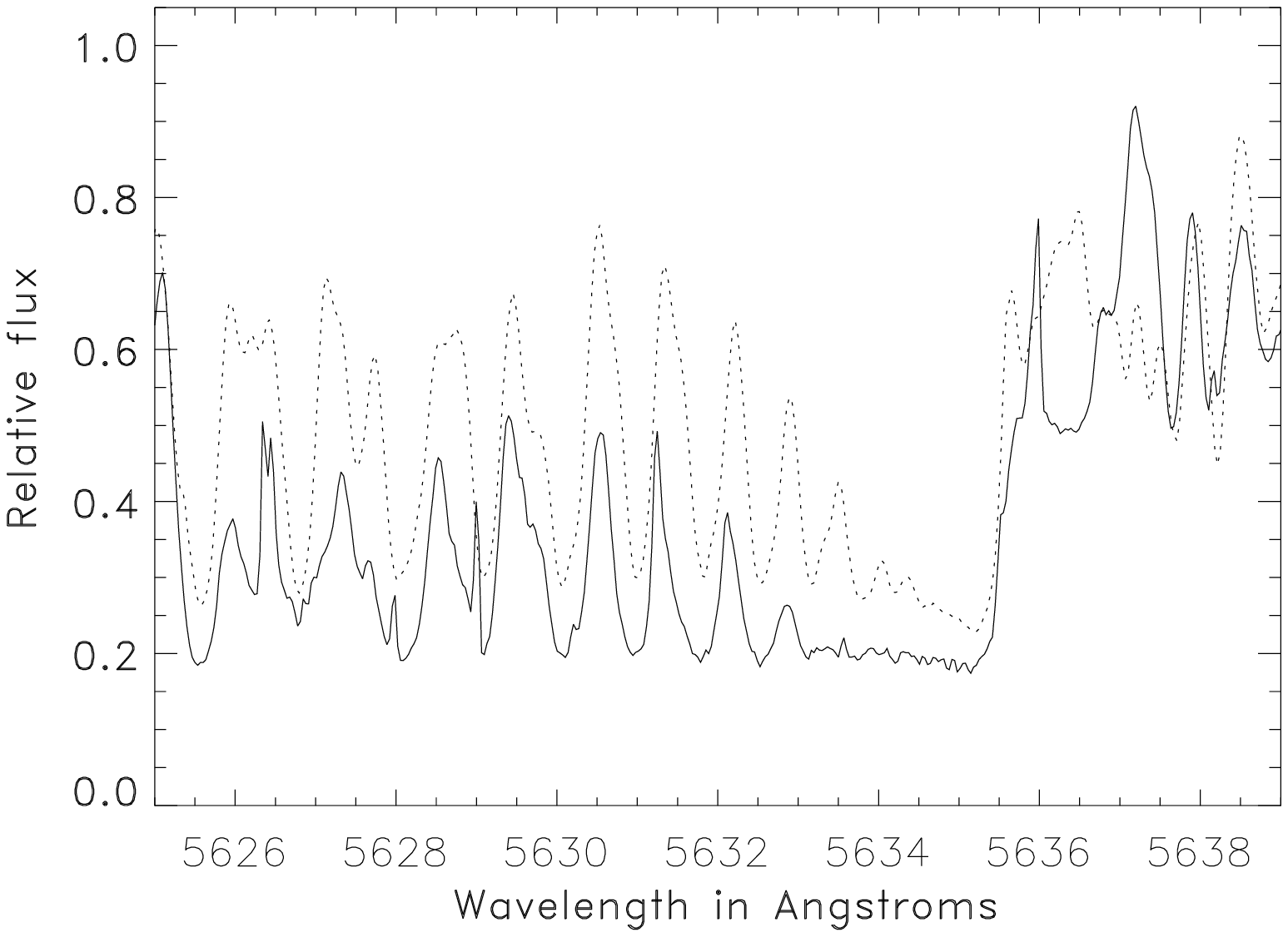,angle=0,width=150mm}}}
\vskip2mm
\captionb{1}{Comparison of the spectra of Z\,UMi (full line) and
HD\,182040 (dotted line) near the C$_2$ (0,1) 563.550\,nm bandhead.}
\vskip2mm
\end{figure}

\newpage

\begin{figure}
\vskip2mm
\centerline{{\psfig{figure=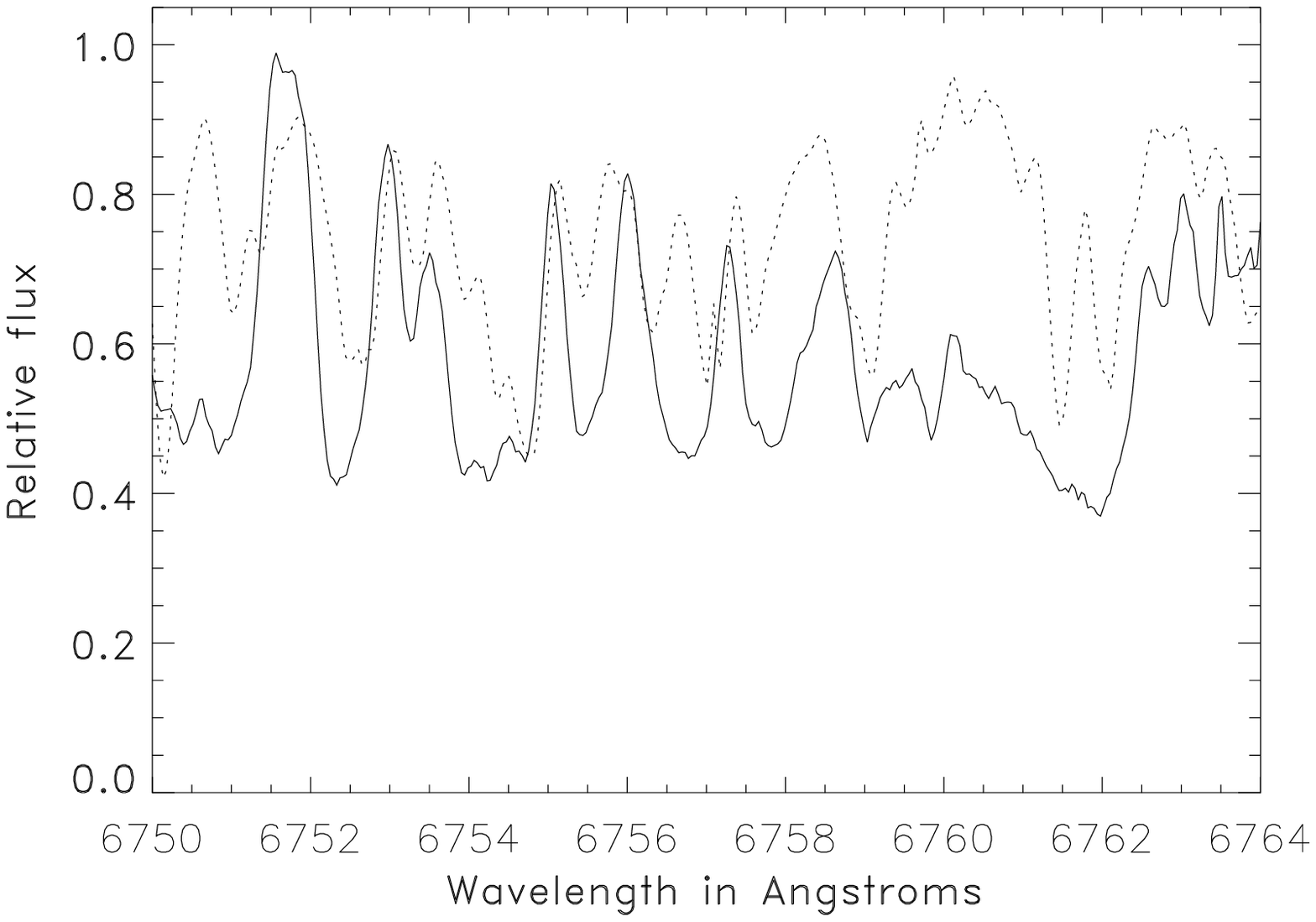,angle=0,width=150mm}}}
\vskip2mm
\captionb{2}{Comparison of the spectra of Z\,UMi (full line) and
HD\,182040 (dotted line) near the C$_2$ (1,4) 676.195\,nm bandhead.}
\vskip2mm
\end{figure}

\newpage

\begin{figure}
\vskip2mm
\centerline{{\psfig{figure=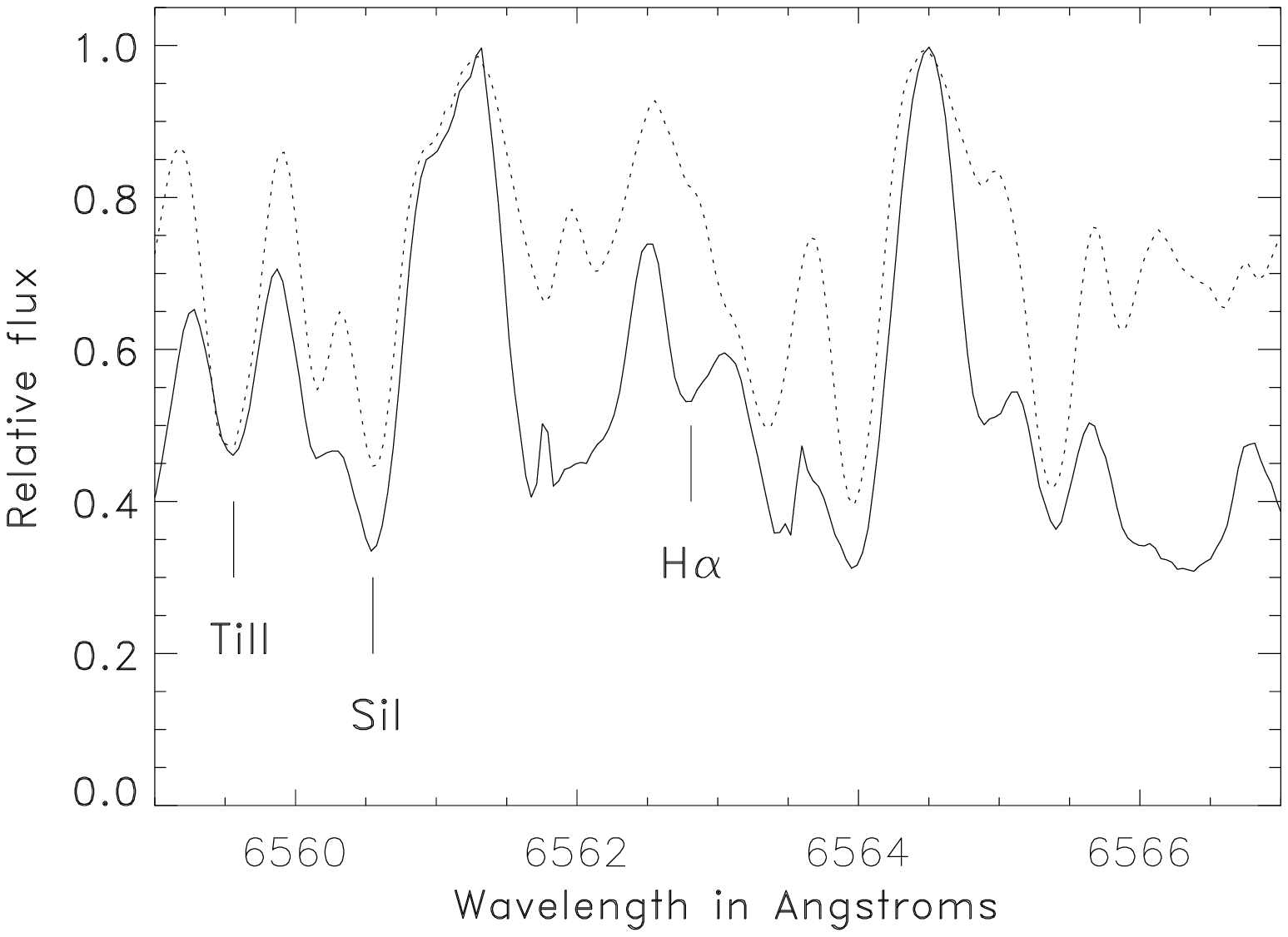,angle=0,width=150mm}}}
\vskip2mm
\captionb{3}{Comparison of the spectra of Z\,UMi (full line) and
HD\,182040 (dotted line) near the H$\alpha$ line. The lines of Ti\,II and
Si\,I are also indicated. All other lines belong to CN.}
\vskip2mm
\end{figure}
                                                                                
\clearpage
\newpage

\begin{center}
\vbox{\footnotesize
\begin{tabular}{lllrrrl}
\multicolumn{7}{c}{\parbox{110mm}{\baselineskip=9pt
{\smallbf Table 1.}{\small\ Chemical composition of Z\,UMi,  of the
majority and minority groups of  R\,CrB stars (Asplund et al. 2000),
and the Sun (normalized to
$\log \Sigma\mu_iA_i=12.15$). The [El/Fe] spread in R\,CrB stars is
given in parentheses. }}}\\[+4pt]
\tablerule
El. & $\log\varepsilon$ & $\log\varepsilon$ &
  & [El/Fe] &
 & Remarks \\
 & Sun$^1$ & Z\,UMi & Z\,UMi & R\,CrB maj. & R\,CrB min. &  \\
\tablerule
H & 12.00 & $\leq10^{-5}$ & & & & \\
Li & 3.25$^2$ & 3.3 & 1.9 & $-0.1(0.8)$ & 0.0(0.5) & \\
C & 8.39 & $9.5\pm0.2$ & & & & C$_2$ bands \\
N & 7.78 & $8.4\pm0.5$ & 2.5 & 1.7(0.3) & 1.9(0.7) & CN bands \\
O & 8.66 & $8.1\pm0.3$ & 1.3 & 0.6(0.6) & 1.5(0.6) & 7$^3$ O\,I and [O\,I] \\
Na& 6.17 &         4.5 & 0.2 & 1.0(0.3) & 1.6(0.2) & 1 Na\,I \\[5pt]
Mg& 7.53 &         5.9 & 0.2 & 0.1(0.3) & $-0.2(0.3)$ & 1 Mg\,I  \\
Al& 6.37 &         5.8 & 1.3 & 0.6(0.3) & 0.9(0.2) & 2 Al\,I  \\
Si& 7.51 & $6.5\pm0.5$ & 0.8 & 0.6(0.3) & 1.6(0.3) & 8 Si\,I  \\
Ca& 6.31 & $3.9\pm0.4$ & $-0.9$ & 0.3(0.2) & 0.2(0.2) & 3 Ca\,I  \\
Sc& 3.05 & $1.5\pm0.5$ & 0.3 & 0.7(0.6) & 1.5(0.1) & 9 Sc\,II  \\[5pt]
Ti& 4.90 & $4.2\pm0.2$ & 1.2 & 0.1(0.4) & 0.5(1.0) & 6 Ti\,I  \\
  &      &  3.2        & 0.2 &          &          & 2 Ti\,II  \\
Cr& 5.64 & $4.0\pm0.3$ & 0.2 &          &          & 7 Cr\,II  \\
Fe& 7.45 & $5.6\pm0.3$ &     &          &          & 77 Fe\,I  \\
  &      &             &     &          &          & 20 Fe\,II  \\[5pt]
Co& 4.92 & $4.2\pm0.3$ & 1.1 &          &          & 4 Co\,I  \\
Ni& 6.25 & $4.9\pm0.3$ & 0.5 & 0.6(0.2) & 1.1(0.3) & 4 Ni\,I  \\       
Y & 2.24 &  0.8        & 0.4 & 0.8(0.6) & 1.2(0.8) & 1 Y\,II  \\
Ba& 2.17 &  0.0        &$-0.3$& 0.3(0.4)& 0.6(0.2) & 1 Ba\,II  \\
La& 1.13 &  0.3        & 1.0 & 1.3(0.5) & 1.1(0.1) & 2 La\,II  \\    
\tablerule
\end{tabular}
}
\end{center}
\centerline{\parbox{110mm}{ \small\baselineskip=9pt
$^1$ Asplund et al. (2005), relative to $\log \varepsilon({\rm H})$,}}
\centerline{\parbox{110mm}{\small\baselineskip=9pt
$^2$ Meteoritic value,}}
\centerline{\parbox{110mm}{\small\baselineskip=9pt
$^3$ Number of used lines.}}

\vskip20mm

\begin{center}
\vbox{\footnotesize
\begin{tabular}{lrrcccc}
\multicolumn{7}{c}{\parbox{110mm}{\baselineskip=9pt
{\smallbf Table 2.}{\small\  The Gaussian decomposition of the Na\,I
       D--lines in the spectrum of Z\,UMi.
      The equivalent widths and FWHMs are given
      in angstroms.}}}\\[+4pt]
\tablerule
No. &$V_{\rm rad}$(D${\rm _2}$) & $V_{\rm rad}$(D${\rm _1}$) & $\rm W_{\lambda}$(D${\rm _2}$) & $\rm W_{\lambda}$(D${\rm _1}$)
            & FWHM(D${\rm _2}$) &  FWHM(D${\rm _1}$)  \\
\tablerule
1&$-$47.0 & $-$43.8 & 0.69 &0.48 & 0.83   & 0.57    \\
2&$-$32.3 & $-$33.0 & 0.07 &0.10 & 0.30   & 0.25    \\
3&$-$6.7  & $-$7.9  & 0.35 &0.32 & 0.41   & 0.37    \\
\tablerule
\end{tabular}
}
\end{center}

\end{document}